\documentclass[useAMS,usenatbib]{mn2e}

\usepackage{graphicx}

\title[Heliospheric Symmetry Explains {\it VOYAGER 2\/}Oobservations]
{Approximate Mirror Symmetry in Heliospheric Plasma Flow Explains {\it VOYAGER 2\/}
Observations}
\author[J. Grygorczuk, A. Czechowski and S. Grzedzielski]
 {J.~Grygorczuk,$^1$\thanks{E-mail:jolagry@cbk.waw.pl (JG); ace@cbk.waw.pl (AC); stangrze@cbk.waw.pl (SG)} 
 A.~Czechowski$^1$\footnotemark[1] and S.~Grzedzielski$^1$\footnotemark[1]\\
  $^1$Space Research Centre PAS, ul.Bartycka 18A, 00-716 Warszawa, Poland}

\begin{document}

\date{Accepted Year Month Day. Received 2015 February Day; in original form 2015 February 11}

\pagerange{\pageref{firstpage}--\pageref{lastpage}} \pubyear{2015}

\maketitle

\label{firstpage}

\begin{abstract}
The Sun and the undisturbed interstellar magnetic field and plasma velocity 
vectors ($\bmath{ B_{IS}}$,$\bmath{ V_{IS}}$) define a mirror symmetry plane 
of the flow at large heliospheric distances.
We show that for the $\bmath {B_{IS}}$ direction defined by {\it IBEX}  
Ribbon center, the radial direction of {\it Voyager 2} over the last decade, 
and the (thermal proton) plasma velocity measured by the spacecraft since 
2010.5, are almost parallel to the  ($\bmath{ B_{IS}}$,$\bmath{ V_{IS}}$)-plane, 
which coincides in practice with the Hydrogen Deflection Plane. These facts  
can be simply explained if approximate mirror symmetry is also maintained 
on the inner side of the heliopause. Such approximate symmetry is possible 
since the solar wind ram pressure is almost spherically symmetric and the 
plasma beta value in the inner heliosheath is high.  In the proposed symmetry, 
the plasma flow speed measured by {\it Voyager 2}  in the inner heliosheath 
is expected to rotate more in the transverse than in the polar direction 
(explanation alternative to \citet{mcsc14}), in evident agreement with 
available spacecraft data (our Fig.1). 
\end{abstract}


\section{Introduction}

Both {\it Voyager} spacecraft continue their journey away from the Sun 
through the heliospheric interface region. {\it Voyager 1} ({\it V1}) 
has already crossed the heliopause (HP) \citep{burl13a,ston13,gurn13}, the boundary 
that separates the 
heliospheric plasma from the interstellar one, and is now operating in 
the interstellar medium at a distance of 130 AU (early 2015) from the 
Sun. {\it Voyager 2} ({\it V2}), at 106.8 AU from the Sun, still 
penetrates the inner heliosheath region.

At the time when {\it Voyagers} commenced their outward journey neither 
the strength nor direction of the interstellar magnetic field (ISMF, $\bmath{ B_{IS}}$) 
was known. In 2009, the {\it Interstellar Boundary Explorer} ({\it IBEX}) 
discovery of the ribbon \citep{mcco09} indicated the probable direction of the ISMF 
as the ribbon center (RC) \citep{funs09}. 

\section[]{Geometrical coincidences}

\begin{figure*} 
\includegraphics[width=170mm]{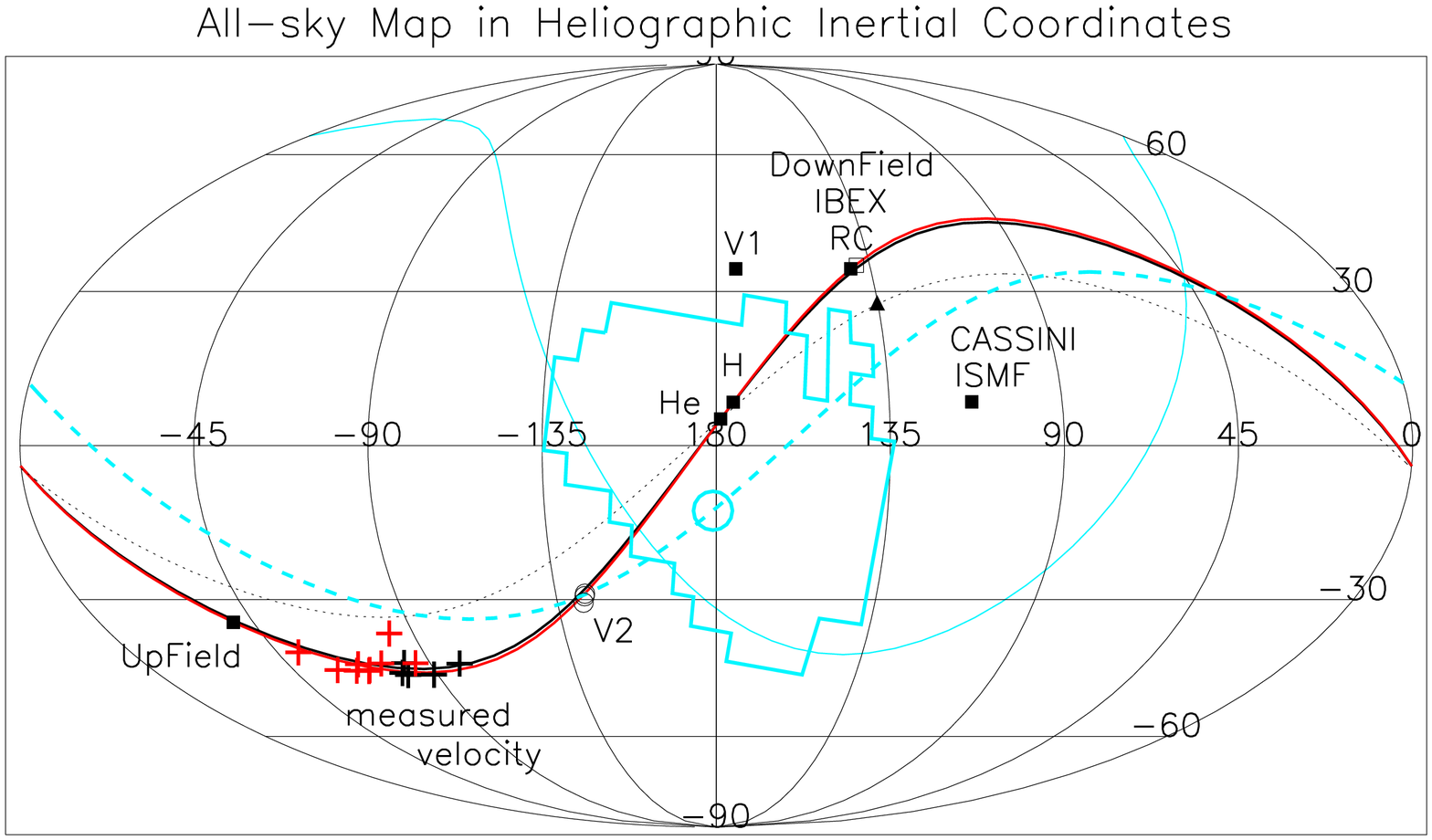}
\caption{Mollweide projection of the plasma velocity directions  
observed by {\it V2} (years 2010 - 2014), as well as {\it V2} locations. 
Shown are the yearly averages of raw data (black crosses) and six-month 
averages of the instrumentally corrected velocity directions (red crosses).
{\it V2} positions in years 2010-2014 are shown by empty circles.  
The black squares indicate the locations of {\it V1} 
in 2012, {\it IBEX} RC (assumed to represent $\bmath{ B_{IS}}$, \citet{funs09}, downfield), 
the direction opposite to RC (upfield),  
the alternative ISMF direction obtained from \mbox{{\it CASSINI}/INCA} measurements 
\citep{dial13}, and the interstellar He \citep{witt04} and H inflow directions. The He 
inflow direction is taken as the direction of the interstellar flow $\bmath{ V_{IS}}$. 
The great circle corresponding to the ($\bmath{ B_{IS}}$,$\bmath{ V_{IS}}$) plane 
(in other words He-RC plane) is shown as a red line, the HDP is black 
(practically coincident with red). 
The dotted line shows the alternative ($\bmath{ B_{IS}}$,$\bmath{ V_{IS}}$) plane 
corresponding to the assumption that $\bmath{ B_{IS}}$ is given by the RC position 
for the highest {\it IBEX} energy (triangle). Blue contours show the high
pressure region and pressure maximum location \citep{schw14,mcsc14}. 
The dashed blue line is the great circle (defined by locations of related 
plasma pressure maximum and the {\it V2} position) on which the crosses corresponding to 
velocity measurements should lie if the flow were organized by pressure 
gradients due to pressure maximum as envisages by \citet{mcsc14}. 
Thin blue line is the approximate location of the {\it IBEX} ribbon.}
\label{f1}
\end{figure*}

It is noteworthy that each of the {\it Voyagers} trajectories happens 
to have a special orientation (instances of geometrical coincidence) 
with respect to the direction defined by the RC. The first such case 
was discussed in \citet{gczg14}. In that paper we pointed out that {\it V1} 
trajectory shares the same heliographic latitude ($\sim$34.5$\degr$) with the RC. 
This led us to a simple explanation why the magnetic field directions 
measured by {\it V1} on both sides of the HP were so similar. 

In the second case, to be discussed now, the {\it V2} spacecraft 
trajectory direction is close (within $\sim$2$\degr$ ) to the plane 
(the ($\bmath{ B_{IS}}$,$\bmath{ V_{IS}}$) plane) which contains the 
position of the Sun and the directions of undisturbed interstellar 
plasma flow ($\bmath{ V_{IS}}$) and the {\it IBEX} RC ($\bmath{ B_{IS}}$). 
This plane was pointed out \citep{gryg11} to be close to the hydrogen 
deflection plane (HDP, the plane defined by the interstellar H and He 
inflow velocity vectors, \citet{lall05,lall10}). 
In current heliospheric modeling, the HDP is generally identified with the 
($\bmath{ B_{IS}}$,$\bmath{ V_{IS}}$) plane 
\citep{izmo05,poza06,ophe06,pogo08,hepo11,katu15}.

If the supersonic solar wind were fully spherically symmetric with no 
interplanetary magnetic field, the ($\bmath{ B_{IS}}$,$\bmath{ V_{IS}}$) 
plane would obviously had been the mirror symmetry plane of the problem. 
The symmetry is broken for a number of causes: fast/slow wind, angle 
and time dependent solar activity, solar rotation, Parker magnetic field, 
etc. However, if these effects are relatively weak, an approximate 
mirror symmetry should be discernible also in the inner heliosheath.

In this letter we show that {\it V2} measurements agree with 
this interpretation. Since {\it V2} position is close to the 
($\bmath{ B_{IS}}$,$\bmath{ V_{IS}}$) plane, 
the plasma velocity field measured by {\it V2} should be approximately 
parallel to this plane. 
We find that this is indeed the case for the plasma flow vectors which come 
from {\it V2} measurements since 2010.5.  

\section{Plasma Velocity Observations}

\begin{figure}
\includegraphics[width=85mm]{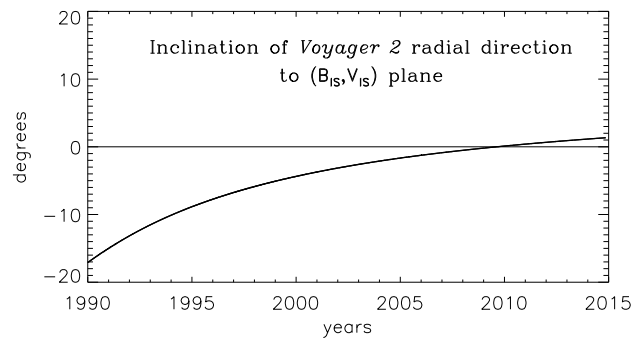}
\caption{Inclination angle of {\it V2} radial direction relative to 
the ($\bmath{ B_{IS}}$,$\bmath{ V_{IS}}$) 
plane versus time.}
\label{f2}
\end{figure}

\begin{figure}
\includegraphics[width=85mm]{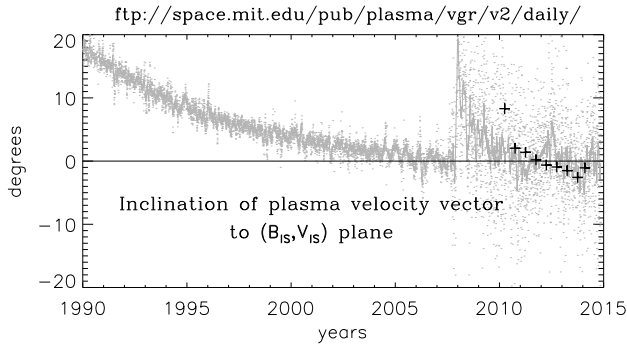}
\caption{The inclination angle of the proton velocity vectors 
to the ($\bmath{ B_{IS}}$,$\bmath{ V_{IS}}$) plane versus time. Shown are daily and 25-day 
averages (dots and line, respectively). Crosses correspond to the angles obtained using 
the corrected plasma velocity vectors \citep{ride14}. The $\bmath{ B_{IS}}$ and $\bmath{ V_{IS}}$ 
are defined as in Figure \ref{f1}.}
\label{f3}
\end{figure}

\begin{figure}
\includegraphics[width=85mm]{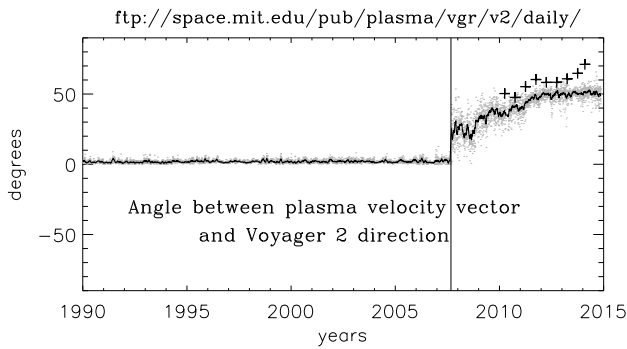}
\caption{The angle between the proton velocity (measured by {\it V2} 
between 1990 and 2015) and the {\it V2} direction. Shown are daily 
and 25-day averages (dots and line, respectively). Crosses correspond 
to the angles obtained using the corrected plasma velocity vectors \citep{ride14}.}  
\label{f4}
\end{figure}

\begin{figure}
\includegraphics[width=85mm]{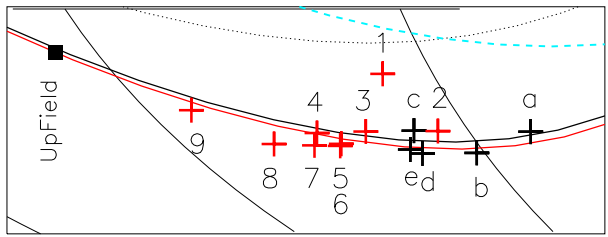}
\caption{Blow-up of part of Figure \ref{f1} showing the evolution of observed plasma 
velocity direction during {\it V2} journey outwards. 
The crosses show the {\it V2} data corrected for the instrument response \citep{ride14} 
(six-month averages, red) and obtained from the raw data (yearly averages, black). 
The data points are numbered (corrected data) or marked by letters consecutively in time. 
The numbers indicate half year periods (1: 2010.25, 2: 2010.75, 3: 2011.25, 
4: 2011.75, 5: 2012.25, 6: 2012.75, 7: 2013.25, 8: 2013.75, 9: 2014.10). The 9th period 
is shorter. The letters correspond to years (a: 2010, b: 2011, c: 2012, d: 2013, e: 2014).}
\label{f5}
\end{figure}
The {\it V2} trajectory inclination relative to the ($\bmath{ B_{IS}}$,$\bmath{ V_{IS}}$) plane has been 
small from \mbox{$\sim$ 2005} (Fig.~\ref{f1}, Fig.~\ref{f2}). At the end of 2014 it is still less than 
2$\degr$   (Fig.~\ref{f2}). Fig.~\ref{f3} shows that, over the last 10 years, excepting abour two  
years after the termination shock crossing by {\it V2}, the plasma flow vectors measured 
by {\it V2} have been approximately parallel to ($\bmath{ B_{IS}}$,$\bmath{ V_{IS}}$) plane, 
a clear confirmation that this plane is indeed an approximate symmetry plane of the flow. 

In Figure \ref{f1}, the {\it V2} positions and measurements  are shown projected on the all-sky map.
We assume that the $\bmath{ B_{IS}}$ direction is the same as the position of the RC according to 
\citet{funs09} (filled square), which is close to the energy-averaged RC position \citep{funs13} 
(empty square). 
The great circles representing the ($\bmath{ B_{IS}}$,$\bmath{ V_{IS}}$) plane (red line) 
and the HDP (black line) are then quite close to each other. 
For measurements between 2010 and 2014, the {\it V2} positions (empty circles) and the yearly averaged 
plasma velocity vectors (black crosses) are shown. The red crosses show the six-month averages of plasma 
velocity direction which was instrumentally corrected \citep{ride14}.

It can be seen how well the measured velocity directions follow the ($\bmath{ B_{IS}}$,$\bmath{ V_{IS}}$) plane
being also close to the HDP. For comparison, we also show (by the dotted line) the position
of the ($\bmath{ B_{IS}}$,$\bmath{ V_{IS}}$) plane for another choice of the $\bmath{ B_{IS}}$ direction
corresponding to the position of RC obtained from the highest energy IBEX data \citep{funs13} 
(triangle). It is clear that the ($\bmath{ B_{IS}}$, $\bmath{ V_{IS}}$) plane defined in this 
way is not a symmetry plane of the flow.    

\section{Discussion}

The restriction of the plasma flow vector at {\it V2} to the ($\bmath{ B_{IS}}$,$\bmath{ V_{IS}}$) plane 
explains also the observation by \citet{mcsc14} that, as {\it V2} moves deeper into the heliosheath, 
the flow speed rotates more in the transverse than in the polar direction. 
In Figure \ref{f1}, this would be represented by the flow direction moving 
along the great circle (representing the ($\bmath{ B_{IS}}$,$\bmath{ V_{IS}}$) plane) 
away from the heliospheric nose and from the {\it V2} trajectory direction 
and consequently towards the region where the circle is approximately tangent 
to the local heliographic parallel. The flow direction would therefore have a substantially 
larger transverse component than the direction of {\it V2} trajectory, 
while staying (approximately) in the same plane as the {\it V2} trajectory and 
$\bmath{ V_{IS}}$ direction.

The symmetry relative to the ($\bmath{ B_{IS}}$,$\bmath{ V_{IS}}$) plane 
(determined solely by the parameters of the local interstellar medium) 
must be to some degree broken by the structure of the inner heliosphere. 
In particular, the solar wind is known to consist of two components 
(slow and fast) with the distribution dependent on the heliographic latitude.
However, the solar wind energy flux (ram pressure) is more symmetric 
\citep{lechat12}, which may reduce the effect on the global heliospheric structure. 

The solar magnetic field (in particular, the Parker spiral) is not symmetric 
relative to the ($\bmath{ B_{IS}}$,$\bmath{ V_{IS}}$) plane. However, its effect on 
the bulk plasma flow in the heliosheath is likely to be small.
It was shown by \citet{mcsc14} based on separation of the {\it IBEX} ENA 
intensities into a part coming from the energetic ions outside the HP 
({\it IBEX} Ribbon) and another part due to ions in the inner heliosheath 
(Global Distributed Flux) \citep{schw14}, that the energy density 
of the heliosheath plasma, determined mainly by the non-thermal ions, 
exceeds by far the magnetic energy density estimated from the {\it V1} 
and {\it V2} data (plasma-beta $\sim$20). This means that the magnetic 
field in the heliosheath can not induce a significant departure from 
the symmetry impressed by the external situation. 

\citet{mcsc14} relate the {\it V2} measurements of the heliospheric plasma
flow to distribution of the integrated plasma pressure inferred from {\it IBEX}
ENA observations \citep{schw14}. They identify the region of high heliosheath
plasma pressure (approximately represented by a blue contour in our Figure \ref{f1})
and argue that the plasma flow is directed away from
the point of maximum pressure (blue circle in our Figure \ref{f1}), which differs 
from the position of the heliospheric nose (He inflow direction). However, we
note that, in this scenario, the flow velocity direction would lie in the 
plane defined by the {\it V2} position and the point of maximum pressure 
(dashed blue line in our Figure \ref{f1}). 

As shown in Figure \ref{f1}, {\it V2} measurements follow the 
($\bmath{ B_{IS}}$,$\bmath{ V_{IS}}$) plane. 
Combination of this (approximate) symmetry with the geometrical coincidence
of {\it V2} position (in years 2013-2015) with the ($\bmath{ B_{IS}}$,$\bmath{ V_{IS}}$) 
plane explains therefore in a natural way the averaged observed plasma velocity
vectors (crosses in Figure \ref{f1}) without recurring to the idea of heliosheath
plasma being pushed away from the excess pressure region near the nose as
speculated in \citet{mcsc14}.

Nevertheless, \citet{mcsc14} idea that the flow 
should be linked with the pressure distribution is important. If a strict mirror
symmetry relative to the ($\bmath{ B_{IS}}$,$\bmath{ V_{IS}}$) plane were valid, one would expect 
that the high plasma pressure region should have the same symmetry. Figure \ref{f1}
shows that the high pressure region is bisected by the ($\bmath{ B_{IS}}$,$\bmath{ V_{IS}}$) 
plane, but not into equal parts. However, the edge of the high pressure region
near the {\it V2} position is approximately perpendicular to the 
($\bmath{ B_{IS}}$,$\bmath{ V_{IS}}$) plane. This implies that the local pressure gradient 
is likely to cause the plasma flow along the plane, as observed by {\it V2}.  

The asymmetry of the pressure distribution obtained by \citet{schw14} and used by 
\citet{mcsc14} is not necessarily in conflict with our results. This is because the
pressure they consider is integrated over the line-of-sight (LOS). It is therefore affected 
also by the region close to the termination shock, where {\it V2} velocity measurements
deviate significantly from the ($\bmath{ B_{IS}}$, $\bmath{ V_{IS}}$) plane (see Figures 
\ref{f3}, \ref{f5}). The velocity measured during first half of the year 2010 is in fact 
close to the dashed blue line in Figures \ref{f1} and \ref{f3}, which, as we have pointed 
out above, corresponds to the pressure distribution used by \citet{mcsc14}. The symmetry
with respect to the ($\bmath{ B_{IS}}$, $\bmath{ V_{IS}}$) may then hold for the local
(not integrated)
pressure distribution at larger distances from the termination shock, where {\it V2}
velocity measurements follow the ($\bmath{ B_{IS}}$, $\bmath{ V_{IS}}$) plane.      

\section{Conclusions}

{\it V2} position near the HDP plane permits a check of mirror symmetry of the heliosheath
(inner and outer) plasma flow with respect to this plane. Our main result is that {\it V2} 
observations are consistent with this symmetry.

Assuming that the $\bmath{ B_{IS}}$ direction is given by the {\it IBEX} RC and  the $\bmath{ V_{IS}}$
direction by the the neutral interstellar He flow, the plasma velocity measured by {\it V2} 
lies in the ($\bmath{ B_{IS}}$,$\bmath{ V_{IS}}$) plane (almost identical with the HDP). 
As {\it V2} approaches the heliopause, the direction of plasma velocity rotates, 
still within the ($\bmath{ B_{IS}}$,$\bmath{ V_{IS}}$) plane. This explains why the transverse 
component of the plasma velocity at {\it V2} becomes larger than the polar one.

The {\it V2} data suggest that the mirror symmetry with respect to ($\bmath{ B_{IS}}$,$\bmath{ V_{IS}}$) plane 
is restricted to the outer part of the inner heliosheath. The plasma pressure 
distribution obtained from {\it IBEX} data is integrated over the full LOS length \citep{schw14}.
This may explain why it departs from this symmetry. 

Because {\it V2} location is approximately in the symmetry plane, we expect that the 
heliospheric plasma flow at the heliopause will be approximately parallel to the 
interstellar plasma flow just outside the heliopause, and anti-parallel to the 
interstellar magnetic field. Figures \ref{f1}, \ref{f4} and \ref{f5} show that the plasma 
velocity at {\it V2} location in fact evolves towards the anti-field direction.

At the heliopause, the draped ISMF would, in general, differ from the undisturbed ISMF. 
However, {\it V2} trajectory direction is close to perpendicular (97$\degr$  ) 
relative to the $\bmath{ B_{IS}}$. In \citet{gczg14} we showed that the magnetic field 
immediately outside the heliopause observed by {\it V2} should then be similar in 
direction to the undisturbed ISMF (RC).

\section*{Acknowledgments}

This work was supported by the Polish National Science Center grant 2012-06-M-ST9-00455.

\bsp

\label{lastpage}

\end{document}